\documentclass[ASNA,twocolumn]{USG}

\usepackage{anyfontsize}
\usepackage{makecell}
\IfFileExists{xurl.sty}{\usepackage{xurl}}{}

\graphicspath{{./}{./images/}}

\articletype{RESEARCH ARTICLE}%

\received{XX Month Year}
\revised{XX Month Year}
\accepted{XX Month Year}
\journal{Journal Name}
\volume{0}
\copyyear{2026}
\startpage{1}
\articledoi{10.1002/advs.XXXXXXX}

\begin{document}

\title{GenMC: Real-Time Generative Monte Carlo Surrogate for Quantitative Photoacoustic Imaging}

\author[1]{Mengjie Shi}
\author[2]{Feng He}
\author[2]{Tom Vercauteren}
\author[2]{Wenfeng Xia}

\authormark{SHI \textsc{et al.}}
\titlemark{GENMC: REAL-TIME GENERATIVE MONTE CARLO SURROGATE FOR QUANTITATIVE PHOTOACOUSTIC IMAGING}

\address[1]{\orgdiv{Department of Bioengineering, }\orgname{Imperial College London, }%
\orgaddress{\city{London }\postcode{SW7 2AZ, }\country{United Kingdom}}}

\address[2]{\orgdiv{School of Biomedical Engineering \& Imaging Sciences, }\orgname{King's College London, }%
\orgaddress{\city{London }\postcode{SE1 7EH, }\country{United Kingdom}}}

\corres{Wenfeng Xia, School of Biomedical Engineering \& Imaging Sciences, King's College London, London SE1 7EH, United Kingdom (\email{wenfeng.xia@kcl.ac.uk})}

\fundingInfo{Wellcome Trust, Grant/Award Number: 203148/Z/16/Z; Engineering and Physical Sciences Research Council, Grant/Award Number: NS/A000049/1}


\abstract[ABSTRACT]{Photoacoustic (PA) imaging provides molecular and functional information about tissue, such as blood oxygen saturation ($sO_2$), yet its clinical translation is hindered by inaccurate quantification. A major source of error is the spectral colouring effect, in which wavelength-dependent optical attenuation distorts the local optical fluence. Monte Carlo (MC) simulation is the gold standard for modelling light transport, but its computational demand precludes real-time use. Here, GenMC is presented, a deep generative framework based on a conditional generative adversarial network that estimates optical fluence distributions from tissue anatomy and literature-derived optical properties, with anatomical priors obtained from co-registered ultrasound images. Trained on MC-generated synthetic datasets, GenMC produces high-fidelity fluence maps in under 30 ms per frame, a four-orders-of-magnitude speed-up over conventional MC simulation, and reaches peak-signal-to-noise ratios of up to 36.24 dB in vivo, outperforming UNet and Pix2Pix baselines. Validation in blood-mimicking phantoms and in 37 human volunteers spanning Fitzpatrick skin types III--V demonstrates improved accuracy, robustness, and physiological consistency of $sO_2$ estimation. By enabling real-time, accurate, and reproducible quantification of tissue oxygenation, GenMC addresses a critical barrier to quantitative PA imaging and offers a general strategy for rapid, high-fidelity approximation of light transport in tissue.}

\copyright{\textcopyright{} 2026 The Authors. \emph{Journal Name} published by Wiley-VCH GmbH.
\\[5pt]
Copyright statement}

\maketitle

\section{Introduction}\label{sec:introduction}

Photoacoustic (PA) imaging has emerged as a powerful modality that combines spectroscopic optical absorption contrast with ultrasound (US) detection, enabling high-resolution imaging at clinically relevant tissue depths \cite{beard2011biomedical, wang2012photoacoustic, ntziachristos2025addressing, omar2019optoacoustic}. By resolving the concentrations of endogenous chromophores, including DNA/RNA, melanin, haemoglobin, lipid, and water, as well as targeted exogenous contrast \cite{weber2016contrast}, quantitative PA imaging offers access to functional and molecular information with considerable potential for disease diagnosis, treatment monitoring, and image-guided interventions \cite{cox2012quantitative, xia2024biomedical, park2025clinical}. A central objective is the non-invasive estimation of blood oxygen saturation ($sO_2$), a key indicator of tissue oxygenation in both physiology and disease. In oncology, $sO_2$ serves as a surrogate marker of tumour hypoxia, a defining feature of malignant progression and therapeutic resistance, making PA imaging a promising approach for cancer diagnosis and treatment monitoring.

Despite these advantages, accurate quantification in PA imaging remains fundamentally constrained by the spectral colouring effect. The measured PA signal is determined not only by optical absorption but also by the spatially varying, wavelength-dependent optical fluence. Whereas absorption encodes biochemical composition, fluence is governed by complex, heterogeneous light transport and cannot be measured directly in vivo. Errors in fluence estimation therefore propagate into substantial inaccuracies in recovered chromophore concentrations and derived functional parameters, including $sO_2$, limiting the reliability and clinical translation of quantitative PA imaging \cite{yuan2006quantitative}.

A range of strategies has been proposed to address this challenge \cite{mondal2025quantitative,tzoumas2016eigenspectra}. Model-based inversion typically involves solving the inverse problem by iteratively minimising the discrepancy between measured PA signals and forward models of light transport, placing critical importance on accurate modelling of the optical fluence distribution. The Diffusion Approximation (DA) to the Radiative Transfer Equation (RTE) is frequently employed due to its computational efficiency and relative simplicity \cite{wang1993hybrid}. However, DA relies on the assumption of isotropic scattering, an approximation that breaks down in regions near the illumination source (within approximately one transport mean free path), and in highly absorbing media where the diffusion regime is not satisfied \cite{jeng2021realtime}. In contrast, Monte Carlo (MC) simulation is widely regarded as the gold standard for modelling light transport in heterogeneous media. By explicitly simulating the stochastic propagation of photon packets, MC provides a rigorous numerical solution to the RTE that remains accurate across diverse optical regimes \cite{wang1995mcml}. However, its high computational cost remains prohibitive: generating high-fidelity fluence maps typically requires simulating $10^7$--$10^9$ photon packets, limiting practical deployment.

Hardware acceleration via graphics processing units (GPUs), such as MCX, has significantly reduced execution times from hours to minutes \cite{fang2009monte}, yet achieving the millisecond-scale latency required for real-time, high-resolution 3D simulations remains challenging. An alternative strategy is to leverage deep learning to accelerate the MC pipeline. For example, denoising frameworks have been developed to transform noise-corrupted fluence maps, generated with low photon counts, into high-quality estimates \cite{raayai2022framework}. However, this level of acceleration remains insufficient for real-time, high-resolution imaging. More recently, data-driven approaches have been explored to bypass traditional numerical solvers altogether \cite{grohl2021deepa}. In particular, deep neural networks have been used to directly retrieve chromophore concentrations from multi-wavelength PA images, often trained on synthetic datasets \cite{cai2018end, bench2020toward}. Although promising, these approaches often exhibit limited domain generalisation to in vivo settings, particularly when trained purely on synthetic data and formulated as direct end-to-end mappings from PA signals to biochemical parameters, thereby restricting their practical utility and clinical translation.

Here, GenMC (Generative MC) is developed, a deep generative framework that redefines how optical fluence is estimated for quantitative PA imaging. Rather than accelerating or approximating conventional photon transport solvers, GenMC replaces them with a learned, data-driven mapping from tissue anatomy derived from co-registered US imaging and optical properties to fluence distributions. The predicted fluence maps are subsequently used to compensate for multi-wavelength PA signals for accurate chromophore unmixing. By eliminating the need for iterative or simulation-based forward modelling, GenMC addresses the fundamental trade-off in quantitative PA imaging between computational efficiency and physical accuracy, where existing methods are typically limited by either high computational cost or reduced fidelity.

\begin{figure*}[!t]
\centerline{\includegraphics[width=0.9\textwidth]{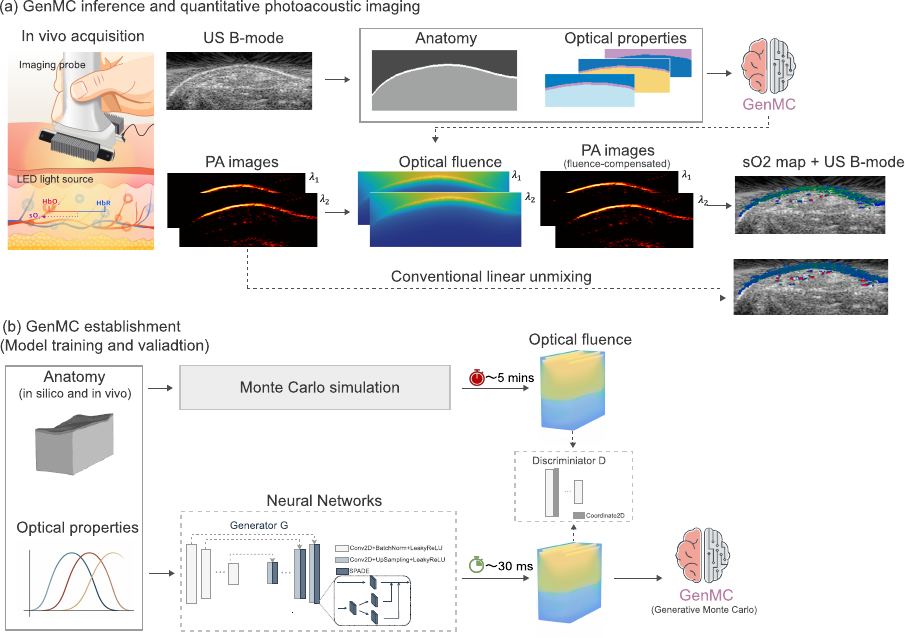}}
\caption{Overview of the Generative Monte Carlo (GenMC) framework for accurate quantitative photoacoustic imaging. a) The inference workflow for in vivo imaging. Following data acquisition using a dual ultrasound (US) and photoacoustic (PA) probe with an LED light source, tissue anatomy is extracted from the US B-mode image. The anatomical map and corresponding tissue optical properties are fed into the trained GenMC model to rapidly predict the spatial distribution of optical fluence at multiple wavelengths ($\lambda_1, \lambda_2$). These predicted fluence maps are utilised to compensate for the raw PA images, enabling the accurate calculation of blood oxygen saturation ($sO_2$). A comparison pipeline demonstrates conventional linear unmixing, which lacks fluence compensation. b) Establishment and training of the GenMC model. Anatomical models and wavelength-dependent optical properties are used to generate ground truth optical fluence maps via conventional Monte Carlo simulation. A neural network architecture comprising a generator (G) utilising spatially adaptive normalisation (SPADE) and a discriminator (D) is trained to synthesise these fluence distributions. The trained GenMC model dramatically accelerates fluence estimation to approximately 30 ms compared to approximately 5 min using Monte Carlo simulation.\label{fig:pipeline}}
\end{figure*}

\section{Results}\label{sec:results}

\subsection{Overview of the GenMC Framework}

An overview of the GenMC framework is presented in Figure~\ref{fig:pipeline}. During training, paired tissue anatomy and wavelength-dependent optical properties are mapped to MC-derived optical fluence distributions using a conditional generative adversarial network (Figure~\ref{fig:pipeline}b). At inference, anatomical priors segmented from co-registered US images are combined with literature-derived optical properties to predict multi-wavelength fluence maps in real time, which are then used to compensate for the measured PA signals before $sO_2$ estimation (Figure~\ref{fig:pipeline}a). The framework was validated in silico, in blood-mimicking phantoms, and in vivo across diverse skin tones; full implementation details are given in the Experimental Section.

\begin{figure*}[!t]
\centerline{\includegraphics[width=0.9\textwidth]{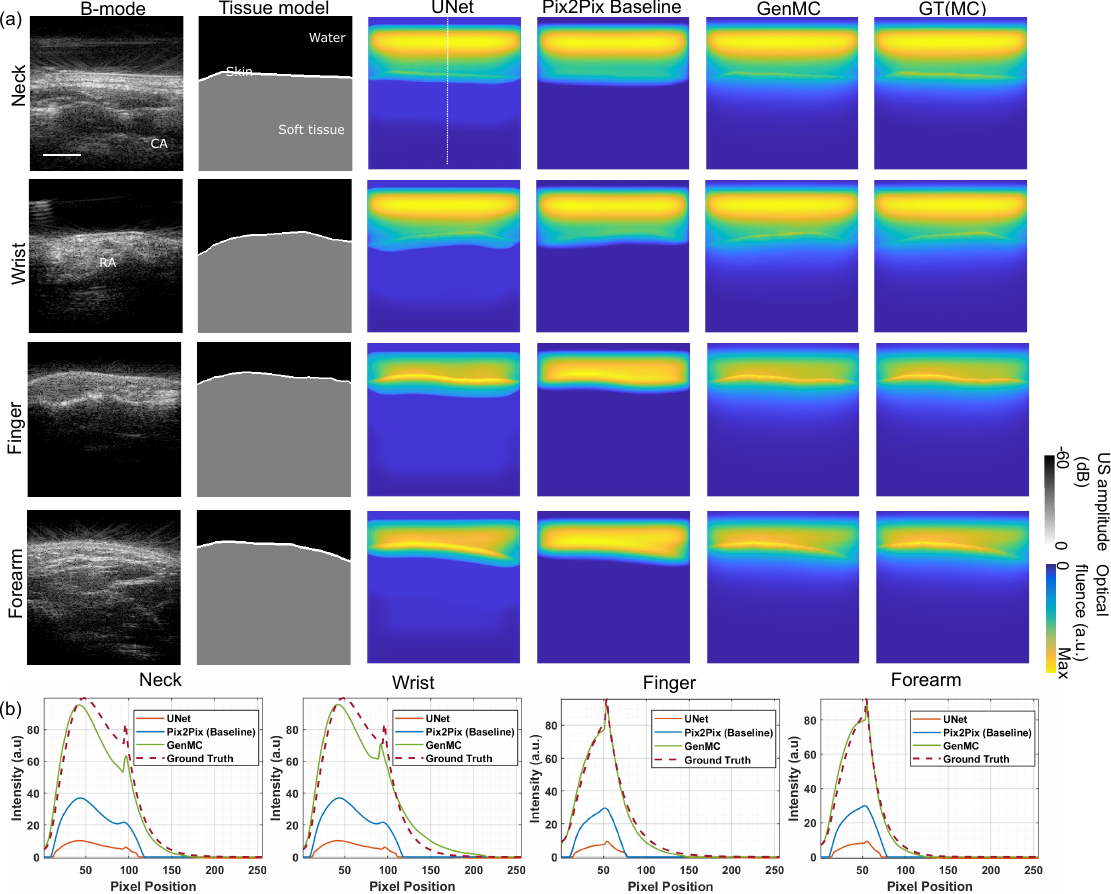}}
\caption{Optical fluence distributions in in vivo human tissues. a) Comparison of GenMC with UNet and Pix2Pix against reference Monte Carlo (MC) simulations. b) Intensity line profiles extracted along the central cross-section (white dashed line). Scale bar: 1 cm. CA: carotid artery; RA: radial artery.\label{results-fig1}}
\end{figure*}

\subsection{Validation of GenMC for Optical Fluence Estimation}

The model was trained exclusively on MC-simulated fluence maps generated from simulated or real tissue geometries with randomly assigned optical properties. Validation was then performed on volunteer-specific anatomical segments derived from in vivo US data that were not represented during training. Performance was compared with established UNet and Pix2Pix baselines (Figure~\ref{results-fig1}). GenMC accurately reconstructed optical fluence distributions across varying tissue geometries, preserving smooth depth-dependent attenuation and continuity across tissue boundaries. In contrast, both UNet and Pix2Pix exhibited pronounced artefacts, particularly at tissue interfaces, where abrupt discontinuities were evident in the cross-sectional profiles (Figure~\ref{results-fig1}b).

Quantitative evaluation was performed using peak signal-to-noise ratio (PSNR), structural similarity index measure (SSIM), and root mean squared error (RMSE) (Table~\ref{tab:overall_metrics}). GenMC achieved an average PSNR of 32.89 dB, exceeding the 30 dB threshold commonly associated with high-fidelity reconstruction. In contrast, UNet and Pix2Pix yielded lower PSNR values of 9.14 dB and 11.80 dB, respectively. Consistently, GenMC achieved substantially higher SSIM and lower RMSE than both baselines, demonstrating its high generalisability and robust accuracy across in vivo conditions.

\begin{table}[!ht]
\centering
\caption{Quantitative performance across 4 anatomical sites each containing 64 measurements (mean $\pm$ standard deviation).\label{tab:overall_metrics}}
\begin{tabular*}{\columnwidth}{@{\extracolsep\fill}lccc@{\extracolsep\fill}}
\toprule
\textbf{Model} & \textbf{PSNR (dB)} & \textbf{SSIM} & \textbf{RMSE ($\%$)} \\
\midrule
UNet & $9.14 \pm 1.58$ & $0.22 \pm 0.05$ & $31.7 \pm 6.9$ \\
Pix2Pix (baseline) & $11.80 \pm 1.58$ & $0.46 \pm 0.04$ & $26.1 \pm 4.6$ \\
GenMC & $\mathbf{32.89 \pm 2.19}$ & $\mathbf{0.93 \pm 0.02}$ & $\mathbf{3.2 \pm 1.3}$ \\
\bottomrule
\end{tabular*}
\begin{tablenotes}
\item PSNR: peak signal-to-noise ratio; SSIM: structural similarity index measure; RMSE: root mean squared error. Best values in bold.
\end{tablenotes}
\end{table}

\subsection{Validation of Quantitative Photoacoustic Imaging in Blood-Mimicking Phantoms}

\begin{figure*}[!t]
\centerline{\includegraphics[width=0.75\textwidth]{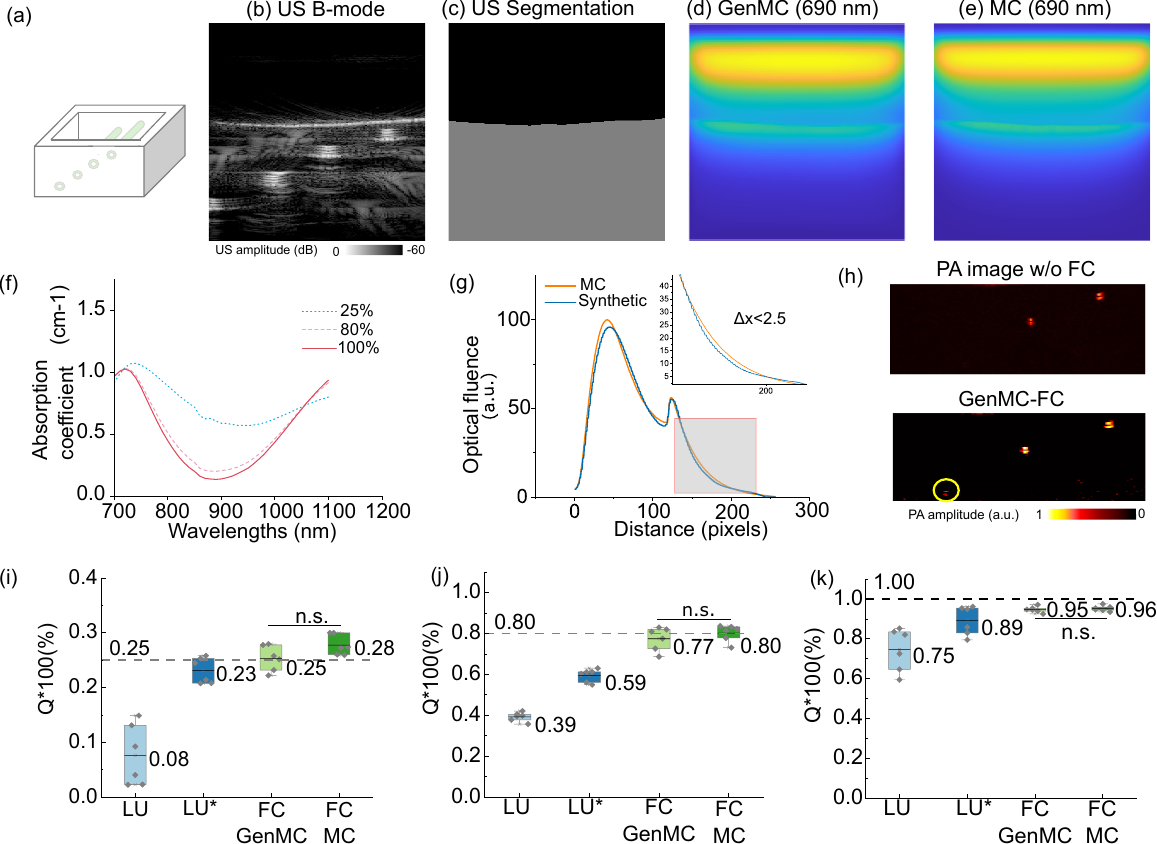}}
\caption{Phantom study evaluating optical fluence estimation and compensation. a--c) A schematic of the experimental phantom, a B-mode ultrasound (US) image, and the corresponding segmentation. d,e) Optical fluence maps at 690 nm generated by d) GenMC and e) Monte Carlo (MC) simulation. f) Optical absorption spectra of the inclusions across wavelengths. g) Depth-resolved 1D fluence profiles extracted from the centre of the imaging plane. h) Photoacoustic (PA) image before (top) and after (bottom) GenMC fluence compensation; the yellow circle indicates the improved visibility of deep targets. i--k) Quantitative assessment of $Q$ (analogy to blood oxygen saturation) across three levels using linear unmixing (LU), linear unmixing with energy normalisation (LU*), GenMC-based fluence compensation (FC-GenMC), and Monte Carlo-based fluence compensation (FC-MC). Each box represents the interquartile range (IQR) of 64 frames obtained from phantoms at given Q levels. The upper and lower edges of the box indicate the 75th and 25th percentiles, respectively, and the central line marks the median. Whiskers extend to the most extreme data points within 1.5 $\times$ IQR of the box boundaries. Black dashed lines indicate the true reference values. Mean values are labelled for each distribution. n.s.: no statistically significant difference.\label{fig-phantom-1}}
\end{figure*}

The absorption coefficients of the phantom inclusions were tuned to mimic three levels of blood oxygen saturation Q (25\%, 80\%, 100\%), based on the mixing ratios reported by Fonseca et~al.\cite{fonseca2017sulfates}. The resulting spectra reproduce the expected signatures reported previously\cite{fonseca2017threedimensionalb} (Figure~\ref{fig-phantom-1}f). Figure~\ref{fig-phantom-1}b,c shows the US image of the phantom and the corresponding segmentation delineating the boundary between the coupling medium (water) and the tissue-mimicking background (1\% Intralipid and 0.0004\% Ink). Figure~\ref{fig-phantom-1}d,e shows the optical fluence distributions generated by GenMC and MC simulation at 690 nm. Line profiles extracted along the central of the imaging plane (Figure~\ref{fig-phantom-1}g) further quantify agreement between the two methods, with a maximum discrepancy of 2.5, demonstrating that GenMC accurately reproduces MC-derived fluence distributions in controlled phantom conditions.

Figure~\ref{fig-phantom-1}h illustrates the depth-dependent attenuation. In the uncorrected image (top), the tube located deeper within the medium is barely visible, whereas applying GenMC-based fluence compensation effectively enhances the signal visibility (yellow circle). The performance of different compensation strategies was investigated using tube phantoms that mimicked different blood oxygen saturation levels (25\%, 80\% and 100\%) (Figure~\ref{fig-phantom-1}i--k). Notably, substantial errors occurred when the pulse energy at different wavelengths was uncorrected (`LU'), and conventional linear unmixing with energy normalisation (`LU*') underestimated Q levels by approximately 2\% to 21\%. Incorporating optical fluence compensation (`FC-GenMC' and `FC-MC') significantly improved accuracy. Statistical analysis revealed no significant difference (n.s.) between GenMC-corrected and MC-corrected Q estimation, demonstrating that GenMC achieves high-fidelity fluence compensation and reliable quantitative Q recovery under controlled phantom conditions.

\subsection{In Vivo Validation of Photoacoustic $sO_2$ Imaging Across Diverse Skin Tones}

\begin{figure*}[!t]
\centerline{\includegraphics[width=0.7\textwidth]{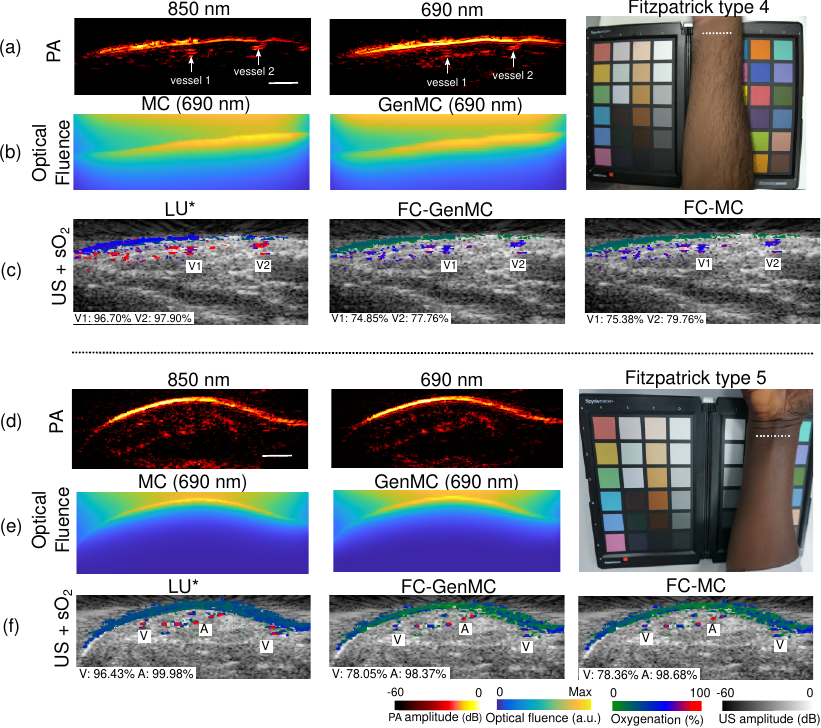}}
\caption{In vivo validation of GenMC-based fluence compensation for blood oxygen saturation ($sO_2$) estimation in subjects with diverse skin tones. Representative results from two subjects are shown. a,d) Photoacoustic (PA) images acquired at 850 nm and 690 nm. b,e) Comparison of optical fluence distributions at 690 nm generated by Monte Carlo (MC) and GenMC. c,f) Ultrasound (US) images overlaid with $sO_2$ maps obtained using linear unmixing with energy normalisation (LU*), GenMC-based fluence compensation (FC-GenMC), and Monte Carlo-based fluence compensation (FC-MC). White dotted lines indicate imaging planes. a.u.: arbitrary unit. Scale bar: 0.5 cm; V: Vein; A: Artery.\label{invivo-1}}
\end{figure*}

To evaluate in vivo performance, GenMC-based fluence compensation was applied for $sO_2$ estimation in human subjects, with two representative cases shown in Figure~\ref{invivo-1}. For the first subject, with Fitzpatrick type IV skin tone, raw PA images at 850 nm and 690 nm clearly delineated two vascular targets (Figure~\ref{invivo-1}a). The optical fluence distribution predicted by GenMC showed close agreement with the MC reference, achieving a PSNR of 26.16 dB (Figure~\ref{invivo-1}b). Without fluence correction, conventional linear unmixing (LU*) resulted in systematic overestimation of $sO_2$, with values exceeding 95\% in both vessels. In contrast, GenMC-based fluence compensation (FC-GenMC) effectively mitigated spectral colouring effects, yielding physiologically plausible venous oxygenation in the range of 74.49\% to 77.93\%. Estimates were in close agreement with those obtained using MC-derived fluence (FC-MC) across all frames, demonstrating both the accuracy and temporal stability of the proposed approach in vivo.

\begin{figure*}[!t]
\centerline{\includegraphics[width=0.7\textwidth]{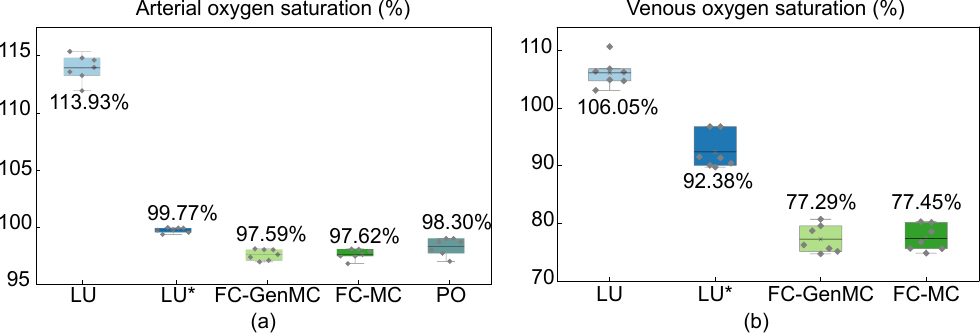}}
\caption{Comparison of arterial a) and venous b) oxygen saturation obtained using different methods. LU: linear unmixing; LU*: conventional linear unmixing with energy normalisation; FC-GenMC: GenMC-based fluence compensation; FC-MC: MC-based fluence compensation; PO: pulse oximetry. Each box represents the interquartile range (IQR) of 64 frames from a single acquisition. The top and bottom edges correspond to the 75th and 25th percentiles, respectively, with central line indicating the median. Whiskers extend to the most extreme data points within 1.5 times the IQR from the box edges.\label{invivo-2}}
\end{figure*}

In the second case, with Fitzpatrick type V skin tone, raw PA images acquired at 850 nm and 690 nm (Figure~\ref{invivo-1}d) exhibited pronounced depth-dependent light attenuation. The optical fluence distribution predicted by the GenMC model at 690 nm closely matched the reference MC simulation (Figure~\ref{invivo-1}e), despite the substantially lower computational cost. The impact of fluence compensation is evident in the overlaid US and $sO_2$ maps (Figure~\ref{invivo-1}f). Without fluence correction, conventional linear unmixing with energy normalisation (LU*) substantially overestimated arterial and venous $sO_2$ (96.43\% and 99.98\%, respectively), whereas FC-GenMC produced physiologically accurate values of 78.05\% (venous) and 98.37\% (arterial), closely matching the MC baseline (78.36\% and 98.68\%) (Figure~\ref{invivo-1}f). Both measurements were additionally validated against simultaneous clinical pulse oximetry, demonstrating agreement within experimental uncertainty.

This performance was further validated through statistical analysis in 15 subjects from a cohort of 37 volunteers (Figure~\ref{invivo-2}). Subjects were included based on a sufficient signal-to-noise ratio to ensure reliable spectral unmixing and quantitative analysis. The cohort consisted of 23 subjects with Fitzpatrick type III, 5 subjects with Fitzpatrick type IV, and 9 subjects with Fitzpatrick type V skin tones. Without fluence correction, conventional linear unmixing (LU) produced non-physical $sO_2$ values exceeding 100\% for both venous ($sO_2$ = 106.05\%) and arterial ($sO_2$ = 113.93\%) measurements. Incorporating energy normalisation (LU*) improved arterial estimates (99.77\%), but continued to overestimate venous $sO_2$ (92.38\%). In contrast, GenMC-based fluence compensation yielded physiologically consistent mean values of 77.29\% (venous) and 97.59\% (arterial), closely matching those obtained using MC-based correction (77.45\% and 97.62\%, respectively). These results demonstrate that GenMC enables accurate and robust $sO_2$ estimation under in vivo conditions, effectively mitigating spectral colouring effects and restoring physiologically meaningful oxygenation measurements.

\section{Discussion}\label{sec:discussion}

Quantitative PA imaging holds great clinical potential for assessing blood $sO_2$. However, its accuracy is fundamentally limited by the spectral colouring effect, which arises from wavelength-dependent optical attenuation and results in spatially and spectrally varying fluence distributions. Consequently, the assumption of wavelength-independent local fluence underlying conventional linear unmixing is violated, leading to inaccurate quantification. While MC simulation remains the gold standard for modelling optical fluence in heterogeneous media, its high computational cost precludes use in real-time clinical workflows. In this study, GenMC bridges this gap by translating the high-fidelity physics of MC simulations into a computationally efficient, end-to-end deep learning framework capable of real-time fluence compensation. This capability is particularly relevant for applications requiring real-time feedback, such as image-guided biopsy, tumour margin assessment, and functional monitoring of tissue oxygenation and haemodynamics, where timely and reliable quantification is essential.

Generative modelling provides a powerful paradigm for learning complex mappings between anatomical structure and underlying physical fields. While such approaches have been successfully applied across a range of biomedical imaging tasks, including retinal fundus synthesis \cite{costa2018endtoendb}, brain magnetic resonance imaging (MRI) \cite{pinaya2022brain}, sparse-view reconstruction in PA tomography \cite{song2023sparse}, and PA image synthesis \cite{rix2023efficient}, their application has largely focused on structural image generation or reconstruction. Even in cases where generative models have been extended to infer physical quantities, such as radiation dose distributions \cite{kearney2020dosegan}, they have not addressed the unique challenges of modelling optical fluence in heterogeneous tissue. In this context, accurate fluence estimation requires not only high spatial fidelity but also strict adherence to the underlying physics of light transport, while operating within the computational constraints of real-time imaging, requirements that existing approaches do not simultaneously satisfy.

GenMC directly addresses this gap by learning a physically grounded mapping from tissue structure and optical properties to fluence distributions. A key strength of this formulation is its ability to preserve the intrinsic relationship between optical properties and light propagation while achieving orders-of-magnitude acceleration in inference. In contrast to model-based inversion approaches, which estimate optical properties by iteratively minimising the discrepancy between measured PA signals and forward models of light transport, and therefore rely on repeated and computationally expensive evaluations of optical fluence, GenMC predicts the fluence distribution directly in a single forward pass. This eliminates the need for iterative optimisation while retaining physical consistency. Furthermore, unlike prior methods that operate either as post hoc denoising within the MC pipeline or as direct end-to-end mappings from PA signals to biochemical parameters, GenMC explicitly models fluence as an intermediate, physically interpretable quantity. This design reduces reliance on implicit correlations and enhances robustness in heterogeneous and previously unseen tissue environments.

GenMC demonstrates robust performance across both controlled phantom experiments and a large in vivo cohort. In blood-mimicking phantoms, the framework effectively corrected the saturation-dependent biases inherent to conventional linear unmixing, particularly at high oxygenation levels where standard approaches substantially underestimate values. In vivo, GenMC was validated across a cohort of 15 human subjects, spanning diverse skin tones, and consistently improved $sO_2$ estimates relative to uncorrected measurements. This included mitigating biases in subjects with darker skin types, where elevated melanin induces strong superficial optical attenuation and acoustic clutter, leading to systematic overestimation of venous and arterial oxygenation in conventional PA oximetry. By incorporating US-derived anatomical information and skin-type-specific optical properties, GenMC accurately models these confounding effects, producing physiologically plausible $sO_2$ estimates while reducing inter-subject variability. Collectively, these findings demonstrate that GenMC provides both mechanistic fidelity and practical generalisability, enabling accurate and equitable quantitative PA imaging across a range of tissue types and optical conditions.

Architecturally, GenMC outperformed conventional supervised baselines, including a standard UNet and a Pix2Pix conditional GAN baseline (PSNR of 32.89 dB versus 9.14 dB and 11.80 dB, respectively). This difference in accuracy stems partly from the output constraints of each framework. While the implemented Pix2Pix terminates in a leakyReLU activation that allows physically impossible negative values, GenMC strictly constrains its output with a ReLU activation. This architectural choice guarantees that fluence predictions remain non-negative and physically admissible. Enforcing non-negativity is a critical correction for the deep, low-fluence regions that dominate the imaging field of view and in which predictions approach zero. Importantly, GenMC addresses a limitation of previous quantitative PA models by explicitly integrating co-registered US anatomical data through SPADE and coordinate layers. Rather than allowing optical properties to be suppressed by standard normalisation, SPADE actively reintroduces this conditioning at every decoder scale. Complementing this, coordinate layers provide the absolute spatial awareness that governs the fluence field. Notably, GenMC is designed to predict the intermediate fluence map rather than directly output $sO_2$, providing a physically interpretable representation of light distribution. This intermediate output is critical for clinical translation, as it allows operators to visually verify predicted fluence before downstream oxygenation calculations, increasing both transparency and confidence in quantitative PA measurements.

Furthermore, GenMC achieves an inference time of under 30 ms on a standard GPU (NVIDIA Tesla T4), enabling real-time fluence compensation in clinical workflows. Beyond PA imaging, this paradigm of accelerated MC simulation has potential applications across a wide range of fields that rely on stochastic transport modelling. For instance, in broader optical modalities, such as diffuse optical tomography (DOT), hyperspectral imaging and tissue spectroscopy, rapid forward modelling of photon migration is essential for feasible iterative image reconstruction and real-time spectral unmixing. The framework's utility extends beyond the optical regime, offering promise for medical physics applications, including radiotherapy treatment planning and clinical dosimetry, where fast, physics-informed simulations can improve both efficiency and accuracy.

Several limitations of the current framework warrant future investigation. First, GenMC currently relies on manual or semi-automatic segmentation of US images to delineate tissue boundaries and assign optical properties. Future iterations could incorporate fully automated, deep learning-based US segmentation models to create an end-to-end pipeline \cite{ma2024segment}. Second, owing to the relatively low pulse energy of the LED light source, only 15 subjects from a cohort of 37 volunteers were included in the quantitative analysis to ensure sufficient signal-to-noise ratio for reliable $sO_2$ estimation. While this reflects a limitation of the current setup, LED-based PA systems offer key advantages for clinical translation, including portability, low cost, and intrinsic eye safety. These features make them well-suited for point-of-care applications, with ongoing improvements in output power and system sensitivity expected to further enhance their utility \cite{singh2020led}. Finally, the current implementation assumes homogeneous optical properties within each segmented layer, assigning uniform, literature-derived values. While this assumption provides a robust baseline that already outperforms conventional linear unmixing, it does not capture the inherent heterogeneity of biological tissues. Crucially, GenMC's millisecond-scale inference and fully differentiable architecture could enable real-time iterative optimisation of optical properties. Internal fluence markers, such as arteries with known oxygen saturation, could guide dynamic calibration via backpropagation, improving fluence estimation and $sO_2$ accuracy \cite{thomas2025quantitative}.

\section{Conclusion}\label{sec:conclusions}

In conclusion, GenMC combines the high-fidelity physics of MC simulations with the speed and flexibility of deep learning, enabling real-time, accurate fluence compensation in both phantom and in vivo studies. By preserving physically interpretable intermediate representations and accounting for diverse skin tones, GenMC advances quantitative PA imaging toward robust, equitable, and clinically translatable applications. Its fast, differentiable architecture not only supports immediate $sO_2$ estimation but also enables future iterative optimisation and adaptive calibration, paving the way for next-generation, real-time, personalised functional imaging. Altogether, GenMC establishes a rapid and reliable framework for quantitative PA imaging, bringing real-time, physiologically meaningful oxygenation measurements closer to widespread clinical adoption.

\section{Experimental Section}\label{sec:experimental}

\subsection{Optical Fluence Synthesis Using Monte Carlo Simulation}\label{Sec.dataset}

Forward light propagation was modelled using Monte Carlo simulations accelerated by MCX \cite{fang2009monte}. A 3-D simulation space of 25.6 (X) $\times$ 27.0 (Y) $\times$ 11.4 (Z) mm was defined, with an isotropic spatial resolution of 0.1 mm. Optical fluence distributions were generated using the MC framework, incorporating LED-based illumination and 3D numerical phantoms.

The light sources were modelled based on a commercial LED-based PA/US imaging probe. Two LED bars were symmetrically arranged with an inter-bar spacing of 1.25 mm. Each bar consisted of a 38 $\times$ 2 array of LED elements (number of elements per row $\times$ number of rows). The emission from each element was modelled as an angular Gaussian beam with a zenith variance of 0.6 voxels, corresponding to an approximate far-field divergence angle of 120 deg. The two LED bars were positioned at an angle of 55 deg with respect to the transducer surface. The spacing between the LED adjacent elements within the array was set to 0.7 mm, while the separation between the two arrays was 1.25 mm. For each MC simulation, a total of $10^{8}$ photons were launched.

Three-dimensional numerical phantoms based on layered tissue models were generated to reflect realistic anatomies and tissue optical properties. Portions of the anatomical structures were generated using three publicly available datasets with tissue annotations \cite{czajkowska2021deep, kronke2022tracked, shi2022improving}. To increase structural diversity, additional variations in the skin layer were introduced using synthetic layered tissue models. These models consisted of either two layers (skin and soft tissues) or three layers (coupling medium, skin, and soft tissues). Variations in the skin surface topology were generated using Gaussian statistics \cite{garcia1984monte}. Moreover, the skin layer thickness was spatially varying, sampled from a discrete uniform distribution $\mathcal{U}_{\mathbb{Z}}\sim(1,9)$ mm.

The optical properties required for MC simulations included the optical absorption coefficient $\mu_a$, reduced optical scattering coefficient $\mu'_s$, anisotropy factor $g$, and refractive index $n$. Parameter values were sampled from the distributions (Table~\ref{tab1}). The optical properties of the skin layer were approximated using a thickness-weighted mean of the epidermis and dermis layers. The epidermis thickness was set in the range of 0.1 to 0.3 mm, while the dermal thickness ranged from 1.3 mm to 2.9 mm.

\begin{table*}[!t]
\centering
\caption{Optical properties of the numerical tissue models.\label{tab1}}
\begin{tabular*}{\textwidth}{@{\extracolsep\fill}ccccc@{\extracolsep\fill}}
\toprule
\textbf{Tissue types} & \textbf{$\mu_a$ [cm$^{-1}$]} & \textbf{$\mu'_s$ [cm$^{-1}$]} & \textbf{$g$} & \textbf{$n$} \\
\midrule
Coupling medium & \makecell[c]{$\mu_{a . \mathrm{water}} \sim \mathcal{U}_{disc}(0.007, 0.07;$ \\ $\Delta = 3\times10 ^{-4})$} & $1^{-10}$ & 1 & 1.33 \\
\midrule
Epidermis & \makecell[c]{$\left[C_M 6.6\left(\lambda^{-3.33}\right)\left(10^{11}\right)\right]+$ \\ $\left(1-C_M\right)\left\{0.244+85.3\left[\exp \left(-\frac{\lambda-154}{66.2}\right)\right]\right\}$}\tnote{$^{\rm a)}$} & $68.7\left(\frac{\lambda}{500}\right)^{-1.16}$ & \makecell[c]{$\mathcal{U}_{disc}(0.80, 0.95;$ \\ $\Delta = 1\times10 ^{-2})$} & 1.44 \\
\midrule
Dermis & \makecell[c]{$C_B \mu_{a . \mathrm{oxy}} + (1-C_B)\times$ \\ $\left(0.244+85.3\left[\exp\left(-\frac{\lambda-154}{66.2}\right)\right]\right)$} & $45.3\left(\frac{\lambda}{500}\right)^{-1.292}$ & \makecell[c]{$\mathcal{U}_{disc}(0.80, 0.95;$ \\ $\Delta = 1\times10 ^{-2})$} & 1.40 \\
\midrule
Soft tissue & \makecell[c]{$C_B C_S \mu_{a . \mathrm{oxy}} + C_B(1-C_S) \mu_{a . \mathrm{deoxy}} +$ \\ $C_W \mu_{a . \mathrm{water}}+ (0.7 - C_W)\mu_{a . \mathrm{fat}}$}\tnote{$^{\rm b),c)}$} & $11.4\left(\frac{\lambda}{500}\right)^{-1.88}$ & \makecell[c]{$\mathcal{U}_{disc}(0.80, 0.90;$ \\ $\Delta = 1\times10 ^{-2})$} & 1.36 \\
\bottomrule
\end{tabular*}
\begin{tablenotes}
\item[$^{\rm a)}$] Wavelength in nm: $\lambda \sim \mathcal{U}_{\mathbb{Z}}(700, 900)$; melanosome fraction: $C_M \sim \mathcal{U}_{disc}(0.06,0.40;\Delta=1\times10^{-2})$, where 6\% corresponds to Caucasian skin and 40\% to pigmented skin.
\item[$^{\rm b)}$] Blood volume fraction: $C_B \sim \mathcal{U}_{disc}(0.002, 0.04; \Delta=1\times10^{-3})$; oxygen saturation fraction: $C_S \sim \mathcal{U}_{disc}(0.6,1.0;\Delta=5\times10^{-2})$; water volume fraction: $C_W \sim \mathcal{U}_{disc}(0.1, 0.7; \Delta= 1 \times 10^{-1})$.
\item[$^{\rm c)}$] $\mu_{a . \mathrm{deoxy}}\sim0.0054*\mathcal{U}_{disc}(772,1794;\Delta=5)$; $\mu_{a . \mathrm{oxy}}\sim 0.0054*\mathcal{U}_{disc}(290,1198;\Delta=5)$; $\mu_{a . \mathrm{fat}}\sim \mathcal{U}_{disc}(0.323,4.633;\Delta= 2 \times 10^{-2})$.
\end{tablenotes}
\end{table*}

A 2D slice corresponding to the imaging plane of the US transducer was extracted from the volumetric optical fluence distribution. The slice was cropped from the superficial region to obtain an image size of 256 $\times$ 256 for model training and validation. Each optical fluence distribution was subsequently normalised to its individual maximum value. A total of 5836 volumetric optical fluence datasets were generated over one week using an NVIDIA Quadro RTX 5000. The source code for the proposed framework is available at \url{https://github.com/MengjieSHI/GenMC-optical-fluence-synthesis}.

\subsection{Network Training}\label{Sec.network}

A deep generative model for optical fluence synthesis was developed based on a cGAN \cite{isola2017image}. The framework comprised a generator G and a discriminator D. The generator G adopts a UNet architecture \cite{ronneberger2015unet}. The encoder block consisted of five downscaling layers, each consisting of two consecutive 3$\times$3 convolution layers followed by ReLU activation and 2$\times$2 max pooling. In the decoder path, each layer concatenated the upsampled feature map from the previous layer with the corresponding encoder features via skip connections, followed by two 3$\times$3 convolution layers and ReLU activation. Batch normalisation in the decoder path was replaced by SPADE \cite{park2019semantic}, enabling the modulation of activations based on spatial context, thereby preserving structural fidelity of the optical fluence distribution. Furthermore, 2D coordinate layers were incorporated to improve the network's spatial awareness, which was critical for modelling the diffusive nature of light propagation.

The discriminator D followed a typical PatchGAN operating on a local scale of 16$\times$16 pixels \cite{isola2017image}. The generator G was trained to estimate the optical fluence distribution based on the corresponding tissue anatomy and optical properties, while the discriminator D was trained to distinguish between the estimated optical fluence distribution $\phi'$ and the corresponding MC-derived ground truth $\phi$.

The loss function $V$ for training a cGAN was a combination of adversarial loss $\mathcal{L}_{cGAN}$ and L1 loss $\mathcal{L}_{L1}$ and expressed as:
\begin{equation}
    V(G, D) = \min_{G}\max_{D} \mathcal{L}_{cGAN}(G, D) + \beta \mathcal{L}_{L1}(G)
    \label{eq:combined}
\end{equation}
where $\mathcal{L}_{cGAN}$ was derived from a binary cross-entropy loss and expressed as:
\begin{equation}
\begin{split}
    \mathcal{L}_{cGAN}(G, D) = \\
    \mathbb{E}_{\phi\sim p(\phi)}[\log(D(\phi))] + \mathbb{E}_{z\sim p(z)}[\log(1-D(G(z)))]
\end{split}
\end{equation}
where $p(\phi)$ denotes the probability distribution of the optical fluence distribution obtained from the MC simulation. $z$ denotes the probability distribution of the optical parameter-encoded tissue segmentation. The G was further regularised using the L1 loss \eqref{eq:l1}.
\begin{equation}
    \mathcal{L}_{L1}(G) = \mathbb{E}_{z, p(\phi)}[\lVert \phi-G(z) \rVert_1]
    \label{eq:l1}
\end{equation}
The weight of L1 regularisation in the combination loss \eqref{eq:combined} was controlled by $\beta$ with an initial value of 10. The network was implemented in Python using PyTorch v1.2.0. The training dataset was randomly split into training, validation, and test sets with a ratio of 8:1:1. Training was performed for 10 epochs with a batch size of 1 using the Adam optimiser (learning rate $2\times10^{-4}$, $\beta_1 = 0.5$, $\beta_2 = 0.999$). The checkpoint achieving the highest validation PSNR was retained. The proposed model was benchmarked against a standard UNet and Pix2Pix baseline using the same optimiser, learning rate, batch size, number of epochs, and data split; the UNet baseline was trained non-adversarially with a mean squared error objective. 

\subsection{Blood Oxygen Saturation Estimation}\label{Sec.so2}

Linear unmixing is the most commonly used method for quantitative PA imaging. The multispectral PA images, corresponding to the initial pressure distributions, can be expressed as:
\begin{equation}
    p_0(\vec{r},\lambda) = \Gamma\mu_a(\vec{r}, \lambda)\phi(\vec{r}, \lambda)
    \label{eq:p0}
\end{equation}
where $\Gamma$ denotes the Gr\"uneisen parameter, $\mu_a(\vec{r},\lambda)$ is the optical absorption coefficient distribution, and $\phi(\vec{r},\lambda)$ is the wavelength-dependent local optical fluence. At wavelengths of 690 nm and 850 nm, optical absorption in tissue is dominated by oxygenated haemoglobin ($HbO_2$) and deoxygenated haemoglobin ($HbR$). Under this assumption, \eqref{eq:p0} can be written as:
\begin{equation}
    p_0(\vec{r},\lambda_i) = \Gamma\phi(\vec{r}, \lambda_i)[c_{HbO_2}(\vec{r})\epsilon_{HbO_2}(\lambda_i)+c_{HbR}(\vec{r})\epsilon_{HbR}(\lambda_i)]
\label{eq-decoloring}
\end{equation}
where $\epsilon_{HbO_2}(\lambda_i)$ and $\epsilon_{HbR}(\lambda_i)$ denote wavelength-dependent molar extinction coefficients of oxygenated and deoxygenated haemoglobin, and $c_{HbO_2}(\vec{r})$ and $c_{HbR}(\vec{r})$ are the corresponding chromophore concentrations.

This formulation can be expressed as a linear system: $Ax = b$, where $A$ is:
\begin{equation}
\begin{aligned}
A&=\begin{bmatrix}
\epsilon_{HbR}(\lambda_1) & \epsilon_{HbO_2}(\lambda_1)\\
\epsilon_{HbR}(\lambda_2) & \epsilon_{HbO_2}(\lambda_2)
\end{bmatrix},\\[0.4em]
x&=\begin{bmatrix} c_{HbR}(\vec{r})\\ c_{HbO_2}(\vec{r})\end{bmatrix},\quad
b=\begin{bmatrix}
\dfrac{p_0(\vec{r},\lambda_1)}{\Gamma\phi(\vec{r},\lambda_1)}\\[0.7em]
\dfrac{p_0(\vec{r},\lambda_2)}{\Gamma\phi(\vec{r},\lambda_2)}
\end{bmatrix}.
\end{aligned}
\label{eq-axb}
\end{equation}
The chromophore concentrations are obtained by solving the linear system:
\begin{equation}
    x = A^{-1}b
    \label{eq-lu}
\end{equation}
from which the blood oxygen saturation $sO_2(\vec{r})$ can be calculated as:
\begin{equation}
    sO_2(\vec{r}) = \frac{c_{HbO_2}(\vec{r})}{c_{HbO_2}(\vec{r})+c_{HbR}(\vec{r})}\times100\%
\end{equation}

In practice, the optical fluence distribution $\phi(\vec{r}, \lambda_i)$ in \eqref{eq-axb} is unknown, and varies spatially and spectrally due to tissue-dependent light attenuation. Conventional linear unmixing methods typically assume $\phi(\vec{r}, \lambda_i)$ to be spatially uniform and identical across wavelengths, such that it can be treated as a constant and cancelled during inversion. Under this assumption, the measured PA signal is directly interpreted as being proportional only to the absorption coefficient $\mu_a(\vec{r},\lambda)$. However, this simplification neglects depth-dependent light attenuation and spectral colouring effects, which introduce substantial bias in the estimated chromophore concentrations and lead to inaccurate $sO_2$ quantification.

In contrast, the GenMC pipeline explicitly accounts for spatially and spectrally varying fluence during spectral unmixing. Rather than assuming a uniform fluence, wavelength-dependent fluence maps $\phi(\vec{r},\lambda)$ are first estimated from the underlying tissue structure using the trained model. These fluence estimates are then incorporated into \eqref{eq-axb} to normalise the measured PA signals before linear inversion. For quantitative evaluation, MC-simulated fluence maps were used as the ground truth.

\subsection{Photoacoustic/Ultrasound Data Acquisition}\label{Sec.acquisition}

Dual-modal PA/US data were acquired using an LED-based PA/US imaging system. The imaging probe consisted of two LED arrays emitting at 690 nm and 850 nm. Each array delivered pulses with an energy of 200 \textmu J, with a pulse repetition frequency of 4 kHz, and a pulse duration of 70 ns. US signals were obtained using a linear array consisting of 128 elements with a central frequency of 7 MHz. Co-registered PA and US images were reconstructed in real time using a Fourier transform-based reconstruction algorithm implemented on a GPU \cite{jaeger2007fourier}. The corresponding raw radio-frequency (RF) data were simultaneously recorded for offline processing.

\subsection{Blood-Mimicking Phantom Study}\label{Sec.phantom}

Three pseudo-blood PA phantoms were constructed for quantitative evaluation. Copper sulphate ($CuSO_4{\cdot}5H_2O$) and nickel sulphate ($NiSO_4{\cdot}6H_2O$) were selected as optical surrogates for $HbO_2$ and $HbR$, respectively \cite{fonseca2017sulfates}. Stock solutions (0.5 M copper sulphate mother solution and 2.2 M nickel sulphate) were mixed according to a ratiometric formulation defined in \eqref{eq:oxygen}, where Q (\%) serves as an analogue of blood oxygen saturation $sO_2$. Absorption spectra of the mixtures were measured using a Vis-NIR spectrophotometer (Agilent, CA, USA). Each phantom comprised four silicone tubes (inner diameter: 0.5 mm) filled with the prepared solutions and embedded in a background medium containing 1\% Intralipid and 0.0004\% India ink to mimic optical scattering and absorption, respectively. Within each phantom, all tubes corresponded to the same Q level (25\%, 80\%, or 100\%). Imaging was conducted with the imaging probe enclosed in a water-filled membrane to ensure acoustic coupling. For each mixture, 1536 B-mode US frames and dual-wavelength PA data (768 frames per wavelength) were collected.
\begin{equation}
    Q(\%) = \frac{\frac{c_{NiSO_4}}{2.2}}{\frac{c_{CuSO_4}}{0.5}+\frac{c_{NiSO_4}}{2.2}}\times 100
    \label{eq:oxygen}
\end{equation}

\subsection{In Vivo Imaging Study}\label{Sec.invivo}

In vivo PA/US data were collected from 37 healthy adult volunteers (BMI 18-25 kg/m$^2$) for model evaluation using the dual-wavelength LED-based PA/US imaging system. The study was approved by the King's College London Research Ethics Committee (study reference: HR-18/19-8881), and all participants provided written informed consent, including consent for publication. Exclusion criteria were participants with active respiratory disorders that may affect blood oxygen saturation levels, a history of cardiovascular disease, medication affecting haemodynamics, use of skin lightening products, tattoos on the imaging areas and a history of skin disorders. Data acquisition comprised three stages: 1) Skin characterisation: the skin tone of the participant was assessed using a colour checker (X-Rite, Calibrite LLC, USA). Photos of the imaged areas were taken under controlled illumination and processed offline to assign a Fitzpatrick skin type. 2) Imaging protocol: a pre-scan was performed at the wrist to localise the radial artery via its pulsation. Dual-wavelength PA imaging was then performed at the same location for approximately 1 min. 3) Reference measurement: during PA/US imaging, peripheral oxygen saturation was recorded using a clinically certified pulse oximeter (Kinetik Wellbeing, Surrey, U.K.) on the middle fingers of the participant at three time points (beginning, middle, and end of the acquisition). Among them, 20 participants were imaged across multiple anatomical sites, including the volar and dorsal forearms, radial and carotid arteries, and fingers for validating GenMC's performance. Owing to the relatively low pulse energy of the LED light source, only 15 subjects from the cohort were included in the quantitative analysis to ensure a sufficient signal-to-noise ratio for reliable $sO_2$ estimation.

For optical fluence estimation, three frames per participant were randomly selected. The upper skin boundary for each selected frame was manually delineated by an experienced operator, and the total skin layer was modelled by extending this boundary to depths of 1.5, 2.0, and 2.5 mm. Skin optical properties were determined based on the Fitzpatrick scale reported in the literature (Table~\ref{tab2}) \cite{setchfield2024effect,tseng2008vivo,tseng2009chromophore}.

\begin{table}[!t]
\centering
\caption{Optical properties at 690 nm and 850 nm.\label{tab2}}
\begin{tabular*}{\columnwidth}{@{\extracolsep\fill}lcccc@{\extracolsep\fill}}
\toprule
\multirow{2}{*}{\textbf{Tissue type}} & \multicolumn{2}{c}{\textbf{$\mu_a$ (mm$^{-1}$)}} & \multicolumn{2}{c}{\textbf{$\mu_s$ (mm$^{-1}$)}} \\
\cmidrule(lr){2-3} \cmidrule(lr){4-5}
 & \textbf{690 nm} & \textbf{850 nm} & \textbf{690 nm} & \textbf{850 nm} \\
\midrule
Coupling medium & 0.0001 & 0.003 & 1.0 & 1.0 \\
Soft tissue & 0.12 & 0.10 & 17.6 & 15.1 \\
\midrule
\multicolumn{5}{l}{\textit{Skin layer}} \\
FST I--II & 0.06 & 0.058 & 20.62 & 15.0 \\
FST III--IV & 0.075 & 0.06 & 21.25 & 16.7 \\
FST V--VI & 0.28 & 0.11 & 18.75 & 11.7 \\
\bottomrule
\end{tabular*}
\begin{tablenotes}
\item FST: Fitzpatrick skin type.
\end{tablenotes}
\end{table}

These parameterised tissue property maps were used as inputs for both conventional MC simulation (reference standard) and the GenMC framework to compute spatially resolved optical fluence distributions. For each volunteer, the three generated fluence maps were averaged to generate one representative GenMC-based map and one MC map. These maps were then used to calculate $sO_2$ across 64 motion-corrected frames.

The $sO_2$ estimation pipeline included energy normalisation followed by spatial fluence compensation utilising either MC- or GenMC-derived fluence maps. To benchmark performance, two baseline methods were considered: 1) the ``LU'' method: conventional linear unmixing without energy normalisation, assuming spatially uniform fluence, and 2) the ``LU*'' method: linear unmixing with energy normalisation and post hoc truncation of $sO_2$ values above 100\% to enforce physiological plausibility.

\subsection{Statistical Analysis}\label{Sec.stats}

\textit{Pre-processing}: For each imaging modality (PA and US), 1536 frames were acquired, with PA data evenly split between the two wavelengths (768 frames per wavelength). To enhance the signal-to-noise ratio, consecutive blocks of 12 frames were averaged, yielding 64 frames per acquisition. Rigid motion correction was subsequently applied to both the PA and US data using the built-in function \textit{imregister} in MATLAB. Each optical fluence distribution was normalised to its individual maximum value, and PA signals were energy-normalised prior to spectral unmixing.

\textit{Data presentation}: Image-quality metrics were computed against the reference MC simulation. For a predicted optical fluence distribution map $\phi'$ and reference MC simulation $\phi$ of size $M{\times}N$, these metrics are defined as:
\begin{equation}
PSNR=10 \log _{10}\left(\frac{\max (\phi)^2}{\frac{1}{M N} \sum_{i=1}^M \sum_{j=1}^N[\phi(i, j)-\phi'(i, j)]^2}\right)
\end{equation}
\begin{equation}
    SSIM = \frac{(2\mu_\phi\mu_{\phi'} + C_1)(2\sigma_{\phi\phi'} + C_2)}{(\mu_\phi^2 + \mu_{\phi'}^2 + C_1)(\sigma_\phi^2 + \sigma_{\phi'}^2 + C_2)}
\end{equation}
where $\mu_\phi$, $\mu_{\phi'}$, $\sigma_\phi$, $\sigma_{\phi'}$, $\sigma_{\phi\phi'}$ denote the mean, standard deviation, and cross-variance of the reference MC simulation $\phi$ and synthetic optical fluence distribution $\phi'$.
\begin{equation}
  \text{RMSE} = \sqrt{\frac{1}{MN} \sum_{i=1}^{M} \sum_{j=1}^{N} \left[ \phi(i,j) - \phi'(i,j) \right]^2}
\end{equation}
Image-quality metrics are reported as mean $\pm$ standard deviation across four anatomical sites, each comprising 64 measurements (Table~\ref{tab:overall_metrics}). Oxygenation estimates are displayed as box plots, in which the box spans the interquartile range (IQR, 25th to 75th percentiles), the central line denotes the median, and the whiskers extend to the most extreme values within 1.5 $\times$ IQR of the box edges.

\textit{Sample size}: $n = 64$ frames per acquisition were analysed for each phantom Q level and for each in vivo acquisition. Phantom experiments comprised three Q levels (25\%, 80\%, 100\%). In vivo, 37 volunteers were recruited, of whom 15 met the signal-to-noise criterion for quantitative $sO_2$ analysis.

\textit{Statistical methods}: Differences between GenMC-based and MC-based fluence-compensated estimates were assessed using paired t-tests, with $p < 0.05$ considered statistically significant; comparisons that did not reach significance are annotated as n.s. in the figures. All analyses were performed in MATLAB R2022b (MathWorks, Natick, MA, USA) and Python (PyTorch v1.2.0).

\bmsubsection*{Supporting Information}
Supporting Information is available from the Wiley Online Library or from the author.

\bmsubsection*{Acknowledgements}
The authors thank Jamie Krens and Hibatulah Adeoye for assistance with volunteer recruitment and in vivo data acquisition, and Dr Virginia Fernandez and Dr Yuxiang Zhou for valuable discussions on model training and high-performance computing. This work was supported in part by the Wellcome Trust [203148/Z/16/Z] and in part by the Engineering and Physical Sciences Research Council (EPSRC) [NS/A000049/1]. For the purpose of Open Access, the authors have applied a CC BY public copyright licence to any Author Accepted Manuscript version arising from this submission.

\bmsubsection*{Conflict of Interest}
T. Vercauteren is a co-founder and shareholder of Hypervision Surgical Ltd., London, U.K. The other authors declare no conflict of interest.

\bmsubsection*{Author Contributions}
M.S.: conceptualization, methodology, software, validation, formal analysis, investigation, data curation, visualization, writing -- original draft. F.H.: data curation, writing -- review and editing. T.V.: supervision, writing -- review and editing. W.X.: conceptualization, resources, supervision, project administration, funding acquisition, writing -- review and editing. All authors read and approved the final manuscript.

\bmsubsection*{Ethics Statement}
The in vivo study was approved by the King's College London Research Ethics Committee (study reference: HR-18/19-8881). All participants provided written informed consent, including consent for publication.

\bmsubsection*{Data Availability Statement}
The source code for the GenMC framework is openly available at \url{https://github.com/MengjieSHI/GenMC-optical-fluence-synthesis}. The in vivo human data are not publicly available owing to participant privacy and the terms of the ethics approval, but may be made available from the corresponding author on reasonable request.

\bibliography{ch5-learning-to-spectral-color}

@article{raayai2022framework,
  title={Framework for denoising Monte Carlo photon transport simulations using deep learning},
  author={Raayai Ardakani, Matin and Yu, Leiming and Kaeli, David R and Fang, Qianqian},
  journal={Journal of Biomedical Optics},
  volume={27},
  number={8},
  pages={083019--083019},
  year={2022},
  publisher={Society of Photo-Optical Instrumentation Engineers}
}

@article{costa2018endtoendb,
  title = {End-to-{{End Adversarial Retinal Image Synthesis}}},
  author = {Costa, Pedro and Galdran, Adrian and Meyer, Maria Ines and Niemeijer, Meindert and Abr{\`a}moff, Michael and Mendon{\c c}a, Ana Maria and Campilho, Aur{\'e}lio},
  year = {2018},
  journal = {IEEE Transactions on Medical Imaging},
  volume = {37},
  number = {3},
  pages = {781--791},
  doi = {10.1109/TMI.2017.2759102}
}

@article{fonseca2017sulfates,
  title = {Sulfates as Chromophores for Multiwavelength Photoacoustic Imaging Phantoms},
  author = {Fonseca, Martina B. and An, Lu and Cox, Benjamin T.},
  year = {2017},
  journal = {Journal of Biomedical Optics},
  volume = {22},
  number = {12},
  pages = {125007},
  publisher = {SPIE},
  doi = {10.1117/1.JBO.22.12.125007},
  urldate = {2024-01-19}
}

@inproceedings{fonseca2017threedimensionalb,
  title = {Three-Dimensional Photoacoustic Imaging and Inversion for Accurate Quantification of Chromophore Distributions},
  booktitle = {{{SPIE BiOS}}},
  author = {Fonseca, Martina and Malone, Emma and Lucka, Felix and Ellwood, Rob and An, Lu and Arridge, Simon and Beard, Paul and Cox, Ben},
  editor = {Oraevsky, Alexander A. and Wang, Lihong V.},
  year = {2017},
  pages = {1006415},
  address = {San Francisco, California, United States},
  doi = {10.1117/12.2250964},
  urldate = {2023-02-18},
  langid = {english}
}

@article{kearney2020dosegan,
  title = {{{DoseGAN}}: A Generative Adversarial Network for Synthetic Dose Prediction Using Attention-Gated Discrimination and Generation},
  author = {Kearney, Vasant and Chan, Jason W. and Wang, Tianqi and Perry, Alan and Descovich, Martina and Morin, Olivier and Yom, Sue S. and Solberg, Timothy D.},
  year = {2020},
  journal = {Scientific Reports},
  volume = {10},
  number = {1},
  pages = {11073},
  publisher = {Nature Publishing Group},
  doi = {10.1038/s41598-020-68062-7},
  urldate = {2024-01-24},
  copyright = {2020 The Author(s)},
  langid = {english}
}

@misc{pinaya2022brain,
  title = {Brain {{Imaging Generation}} with {{Latent Diffusion Models}}},
  author = {Pinaya, Walter H. L. and Tudosiu, Petru-Daniel and Dafflon, Jessica and {da Costa}, Pedro F. and Fernandez, Virginia and Nachev, Parashkev and Ourselin, Sebastien and Cardoso, M. Jorge},
  year = {2022},
  number = {arXiv:2209.07162},
  eprint = {2209.07162},
  primaryclass = {cs, eess, q-bio},
  publisher = {arXiv},
  urldate = {2022-11-30},
  archiveprefix = {arxiv},
  langid = {english}
}

@article{tseng2009chromophore,
  title = {Chromophore Concentrations, Absorption and Scattering Properties of Human Skin in-Vivo},
  author = {Tseng, Sheng-Hao and Bargo, Paulo and Durkin, Anthony and Kollias, Nikiforos},
  year = {2009},
  journal = {Optics Express},
  volume = {17},
  number = {17},
  pages = {14599},
  doi = {10.1364/OE.17.014599},
  urldate = {2023-07-13},
  langid = {english}
}

@article{wang1993hybrid,
  title = {Hybrid Model of {{Monte Carlo}} Simulation and Diffusion Theory for Light Reflectance by Turbid Media},
  author = {Wang, Lihong and Jacques, Steven L.},
  year = {1993},
  journal = {JOSA A},
  volume = {10},
  number = {8},
  pages = {1746--1752},
  publisher = {Optica Publishing Group},
  doi = {10.1364/JOSAA.10.001746},
  urldate = {2024-01-25},
  copyright = {{\copyright} 1993 Optical Society of America},
  langid = {english}
}

@inproceedings{isola2017image,
  title={Image-to-image translation with conditional adversarial networks},
  author={Isola, Phillip and Zhu, Jun-Yan and Zhou, Tinghui and Efros, Alexei A},
  booktitle={Proceedings of the IEEE conference on computer vision and pattern recognition},
  pages={1125--1134},
  year={2017}
}

@inproceedings{park2019semantic,
  title={Semantic image synthesis with spatially-adaptive normalization},
  author={Park, Taesung and Liu, Ming-Yu and Wang, Ting-Chun and Zhu, Jun-Yan},
  booktitle={Proceedings of the IEEE/CVF conference on computer vision and pattern recognition},
  pages={2337--2346},
  year={2019}
}

@article{beard2011biomedical,
  title = {Biomedical Photoacoustic Imaging},
  author = {Beard, Paul},
  year = {2011},
  journal = {Interface Focus},
  volume = {1},
  number = {4},
  pages = {602--631},
  doi = {10.1098/rsfs.2011.0028},
  langid = {english},
  pmcid = {PMC3262268},
  pmid = {22866233}
}

@article{fang2009monte,
  title = {Monte {{Carlo Simulation}} of {{Photon Migration}} in {{3D Turbid Media Accelerated}} by {{Graphics Processing Units}}},
  author = {Fang, Qianqian and Boas, David A.},
  year = {2009},
  journal = {Opt. Express, OE},
  volume = {17},
  number = {22},
  pages = {20178--20190},
  publisher = {{Optical Society of America}},
  doi = {10.1364/OE.17.020178},
  copyright = {\&\#169; 2009 OSA},
  langid = {english}
}

@article{jaeger2007fourier,
  title = {Fourier Reconstruction in Optoacoustic Imaging Using Truncated Regularized Inverse k -Space Interpolation},
  author = {Jaeger, Michael and Sch{\"u}pbach, Simon and Gertsch, Andreas and Kitz, Michael and Frenz, Martin},
  year = {2007},
  journal = {Inverse Problems},
  volume = {23},
  number = {6},
  pages = {S51--S63},
  publisher = {{IOP Publishing}},
  doi = {10.1088/0266-5611/23/6/S05},
  langid = {english}
}

@article{wang1995mcml,
  title = {{{MCML}}\textemdash{{Monte Carlo}} Modeling of Light Transport in Multi-Layered Tissues},
  author = {Wang, Lihong and Jacques, Steven L. and Zheng, Liqiong},
  year = {1995},
  journal = {Computer Methods and Programs in Biomedicine},
  volume = {47},
  number = {2},
  pages = {131--146},
  doi = {10.1016/0169-2607(95)01640-F},
  langid = {english}
}

@article{grohl2021deepa,
  title = {Deep Learning for Biomedical Photoacoustic Imaging: {{A}} Review},
  author = {Gr{\"o}hl, Janek and Schellenberg, Melanie and Dreher, Kris and {Maier-Hein}, Lena},
  year = {2021},
  journal = {Photoacoustics},
  volume = {22},
  pages = {100241},
  doi = {10.1016/j.pacs.2021.100241},
  urldate = {2021-12-05},
  langid = {english}
}

@article{jeng2021realtime,
  title = {Real-Time Interleaved Spectroscopic Photoacoustic and Ultrasound ({{PAUS}}) Scanning with Simultaneous Fluence Compensation and Motion Correction},
  author = {Jeng, Geng-Shi and Li, Meng-Lin and Kim, MinWoo and Yoon, Soon Joon and Pitre, John J. and Li, David S. and Pelivanov, Ivan and O'Donnell, Matthew},
  year = {2021},
  journal = {Nature Communications},
  volume = {12},
  number = {1},
  pages = {716},
  publisher = {Nature Publishing Group},
  doi = {10.1038/s41467-021-20947-5},
  urldate = {2023-02-16},
  copyright = {2021 The Author(s)},
  langid = {english}
}

@article{garcia1984monte,
  title={Monte Carlo calculation for electromagnetic-wave scattering from random rough surfaces},
  author={Garcia, N and Stoll, E},
  journal={Physical review letters},
  volume={52},
  number={20},
  pages={1798},
  year={1984},
  publisher={APS}
}

@article{ronneberger2015unet,
  title = {U-{{Net}}: {{Convolutional Networks}} for {{Biomedical Image Segmentation}}},
  author = {Ronneberger, Olaf and Fischer, Philipp and Brox, Thomas},
  year = {2015},
  journal = {arXiv:1505.04597 [cs]},
  eprint = {1505.04597},
  primaryclass = {cs},
  urldate = {2021-12-06},
  archiveprefix = {arxiv}
}

@article{cox2012quantitative,
  title={Quantitative spectroscopic photoacoustic imaging: a review},
  author={Cox, Ben and Laufer, Jan G and Arridge, Simon R and Beard, Paul C},
  journal={Journal of biomedical optics},
  volume={17},
  number={6},
  pages={061202--061202},
  year={2012},
  publisher={Society of Photo-Optical Instrumentation Engineers}
}

@article{setchfield2024effect,
  title={Effect of skin color on optical properties and the implications for medical optical technologies: a review},
  author={Setchfield, Kerry and Gorman, Alistair and Simpson, A Hamish RW and Somekh, Michael G and Wright, Amanda J},
  journal={Journal of biomedical optics},
  volume={29},
  number={1},
  pages={010901--010901},
  year={2024},
  publisher={Society of Photo-Optical Instrumentation Engineers}
}

@article{tseng2008vivo,
  title={In vivo determination of skin near-infrared optical properties using diffuse optical spectroscopy},
  author={Tseng, Sheng-Hao and Grant, Alexander and Durkin, Anthony J},
  journal={Journal of biomedical optics},
  volume={13},
  number={1},
  pages={014016--014016},
  year={2008},
  publisher={Society of Photo-Optical Instrumentation Engineers}
}

@article{bench2020toward,
  title={Toward accurate quantitative photoacoustic imaging: learning vascular blood oxygen saturation in three dimensions},
  author={Bench, Ciaran and Hauptmann, Andreas and Cox, Ben},
  journal={Journal of Biomedical Optics},
  volume={25},
  number={8},
  pages={085003--085003},
  year={2020},
  publisher={Society of Photo-Optical Instrumentation Engineers}
}

@article{cai2018end,
  title={End-to-end deep neural network for optical inversion in quantitative photoacoustic imaging},
  author={Cai, Chuangjian and Deng, Kexin and Ma, Cheng and Luo, Jianwen},
  journal={Optics letters},
  volume={43},
  number={12},
  pages={2752--2755},
  year={2018},
  publisher={Optical Society of America}
}

@article{song2023sparse,
  title={Sparse-view reconstruction for photoacoustic tomography combining diffusion model with model-based iteration},
  author={Song, Xianlin and Wang, Guijun and Zhong, Wenhua and Guo, Kangjun and Li, Zilong and Liu, Xuan and Dong, Jiaqing and Liu, Qiegen},
  journal={Photoacoustics},
  volume={33},
  pages={100558},
  year={2023},
  publisher={Elsevier}
}

@article{ma2024segment,
  title={Segment anything in medical images},
  author={Ma, Jun and He, Yuting and Li, Feifei and Han, Lin and You, Chenyu and Wang, Bo},
  journal={Nature communications},
  volume={15},
  number={1},
  pages={654},
  year={2024},
  publisher={Nature Publishing Group UK London}
}

@article{czajkowska2021deep,
  title={Deep learning approach to skin layers segmentation in inflammatory dermatoses},
  author={Czajkowska, Joanna and Badura, Pawel and Korzekwa, Szymon and P{\l}atkowska-Szczerek, Anna},
  journal={Ultrasonics},
  volume={114},
  pages={106412},
  year={2021},
  publisher={Elsevier}
}

@article{kronke2022tracked,
  title={Tracked 3D ultrasound and deep neural network-based thyroid segmentation reduce interobserver variability in thyroid volumetry},
  author={Kr{\"o}nke, Markus and Eilers, Christine and Dimova, Desislava and K{\"o}hler, Melanie and Buschner, Gabriel and Schweiger, Lilit and Konstantinidou, Lemonia and Makowski, Marcus and Nagarajah, James and Navab, Nassir and others},
  journal={Plos one},
  volume={17},
  number={7},
  pages={e0268550},
  year={2022},
  publisher={Public Library of Science San Francisco, CA USA}
}

@article{shi2022improving,
  title={Improving needle visibility in LED-based photoacoustic imaging using deep learning with semi-synthetic datasets},
  author={Shi, Mengjie and Zhao, Tianrui and West, Simeon J and Desjardins, Adrien E and Vercauteren, Tom and Xia, Wenfeng},
  journal={Photoacoustics},
  volume={26},
  pages={100351},
  year={2022},
  publisher={Elsevier}
}

@article{thomas2025quantitative,
  title={Quantitative photoacoustic imaging using known chromophores as fluence marker},
  author={Thomas, Anjali and Rietberg, Max and Akkus, Mervenur and van Soest, Gijs and Francis, Kalloor Joseph},
  journal={Photoacoustics},
  volume={41},
  pages={100673},
  year={2025},
  publisher={Elsevier}
}

@article{rix2023efficient,
  title={Efficient photoacoustic image synthesis with deep learning},
  author={Rix, Tom and Dreher, Kris K and N{\"o}lke, Jan-Hinrich and Schellenberg, Melanie and Tizabi, Minu D and Seitel, Alexander and Maier-Hein, Lena},
  journal={Sensors},
  volume={23},
  number={16},
  pages={7085},
  year={2023},
  publisher={MDPI}
}

@article{wang2012photoacoustic,
  title={Photoacoustic tomography: in vivo imaging from organelles to organs},
  author={Wang, Lihong V and Hu, Song},
  journal={science},
  volume={335},
  number={6075},
  pages={1458--1462},
  year={2012},
  publisher={American Association for the Advancement of Science}
}

@article{ntziachristos2025addressing,
  title={Addressing unmet clinical need with optoacoustic imaging},
  author={Ntziachristos, Vasilis},
  journal={Nature Reviews Bioengineering},
  volume={3},
  number={3},
  pages={182--184},
  year={2025},
  publisher={Nature Publishing Group UK London}
}

@article{omar2019optoacoustic,
  title={Optoacoustic mesoscopy for biomedicine},
  author={Omar, Murad and Aguirre, Juan and Ntziachristos, Vasilis},
  journal={Nature biomedical engineering},
  volume={3},
  number={5},
  pages={354--370},
  year={2019},
  publisher={Nature Publishing Group UK London}
}

@article{weber2016contrast,
  title={Contrast agents for molecular photoacoustic imaging},
  author={Weber, Judith and Beard, Paul C and Bohndiek, Sarah E},
  journal={Nature methods},
  volume={13},
  number={8},
  pages={639--650},
  year={2016},
  publisher={Nature Publishing Group US New York}
}

@book{xia2024biomedical,
  title={Biomedical Photoacoustics: technology and applications},
  author={Xia, Wenfeng},
  year={2024},
  publisher={Springer}
}

@article{park2025clinical,
  title={Clinical translation of photoacoustic imaging},
  author={Park, Jeongwoo and Choi, Seongwook and Knieling, Ferdinand and Clingman, Bryan and Bohndiek, Sarah and Wang, Lihong V and Kim, Chulhong},
  journal={Nature Reviews Bioengineering},
  volume={3},
  number={3},
  pages={193--212},
  year={2025},
  publisher={Nature Publishing Group UK London}
}

@article{yuan2006quantitative,
  title={Quantitative photoacoustic tomography: Recovery of optical absorption coefficient maps of heterogeneous media},
  author={Yuan, Zhen and Jiang, Huabei},
  journal={Applied physics letters},
  volume={88},
  number={23},
  year={2006},
  publisher={AIP Publishing}
}

@article{mondal2025quantitative,
  title={Quantitative photoacoustic tomography: imaging tissue properties and functional parameters accurately},
  author={Mondal, Sudeep and Jiang, Huabei},
  journal={Advanced Imaging},
  volume={22002},
  pages={1},
  year={2025}
}

@article{tzoumas2016eigenspectra,
  title={Eigenspectra optoacoustic tomography achieves quantitative blood oxygenation imaging deep in tissues},
  author={Tzoumas, Stratis and Nunes, Antonio and Olefir, Ivan and Stangl, Stefan and Symvoulidis, Panagiotis and Glasl, Sarah and Bayer, Christine and Multhoff, Gabriele and Ntziachristos, Vasilis},
  journal={Nature communications},
  volume={7},
  number={1},
  pages={12121},
  year={2016},
  publisher={Nature Publishing Group UK London}
}

@article{singh2020led,
  title={Led-based photoacoustic imaging},
  author={Singh, M Kuniyil Ajith and Sato, N and Ichihashi, F and Sankai, Y},
  journal={Springer},
  volume={10},
  pages={978--981},
  year={2020},
  publisher={Springer}
}

\end{document}